\documentclass{article}
\usepackage{glas}
\usepackage{epsf}
\epsfverbosetrue
\def\Journal#1#2#3#4{{#1} {\bf #2}, (#4) #3}
\def\PRD{{\em Phys. Rev.} D}

\def\PLB{{\em Phys. Lett.}  B}
\def\NPB{{\em Nucl. Phys.} B}
\def\CPC{\em Comp. Phys. Comm.}
\def\oos{${\cal O}(\alpha \alpha_s^2)$}
\begin{document}
\begin{titlepage}{GLAS-PPE/98--01}{May 1998}
\title{Dynamics of Multijet Processes in Photoproduction}
\author{L.\ E.\ Sinclair}
\collaboration{for the ZEUS Collaboration}
\begin{abstract}
The cross section for producing three high transverse energy jets 
with a three-jet invariant mass of $M_{3J} > 50$~GeV has been
measured in photoproduction at HERA.  The angular distribution of 
the three jets is presented and found to be inconsistent with a
uniform population of the available phase space.  In contrast, 
parton shower models and \oos\ pQCD 
calculations are able to describe the three jet dynamics.
\end{abstract}
\vfill
\conference{talk given at the 6th International Workshop \\on Deep 
Inelastic Scattering and QCD (DIS '98), \\
Brussels, Belgium. \\April 4-8, 1998}
\end{titlepage}
\section{Introduction}

The study of multijet production provides a direct test of perturbative
QCD (pQCD) predictions beyond leading order.  It provides in addition  
sensitive tests of extensions to fixed order theories such as parton
shower models.  We have measured distributions of observables which span
the multijet parameter space and have a relatively straightforward 
interpretation within pQCD~\cite{Geer_96}.  Such 
distributions have previously been  measured for two jet production at 
HERA~\cite{ZEUS_96} and for three or more jets at the 
Tevatron~\cite{D0_96,CDF_96}.  Here these distributions are presented
for inclusive three-jet photoproduction for the first time.

The measured quantities are the three-jet invariant mass, $M_{3J}$, and
the dimensionless quantities $X_3$, $X_4$, $\cos \theta_3$ and $\psi_3$
which determine the configuration of the three jets.
These quantities are defined,
\begin{displaymath}
X_3 \equiv \frac{2E_3}{M_{3J}},\  X_4 \equiv \frac{2E_4}{M_{3J}},\ 
\cos \theta_3 \equiv \frac{\vec{p}_{AV} \cdot \vec{p}_3}{|\vec{p}_{AV}| |\vec{p}_3|}\  \mbox{and} \  
\cos{\psi_3} \equiv \frac{(\vec{p}_3 \times \vec{p}_{AV}) \cdot 
                              (\vec{p}_4 \times \vec{p}_5)}
                              {|\vec{p}_3 \times \vec{p}_{AV}| 
                               |\vec{p}_4 \times \vec{p}_5|},
\end{displaymath}
where $E_j$ and $\vec{p}_j$ are the energies and three-vectors respectively
of the jets in this frame, the jets are numbered 3, 4 and 5 in order
of energy decreasing and $\vec{p}_{AV}$ represents the average 
beam direction.

\begin{figure}[htb]
\begin{flushright}
\leavevmode
\makebox[8.cm][l]{\epsfxsize=5.0cm.\epsfbox{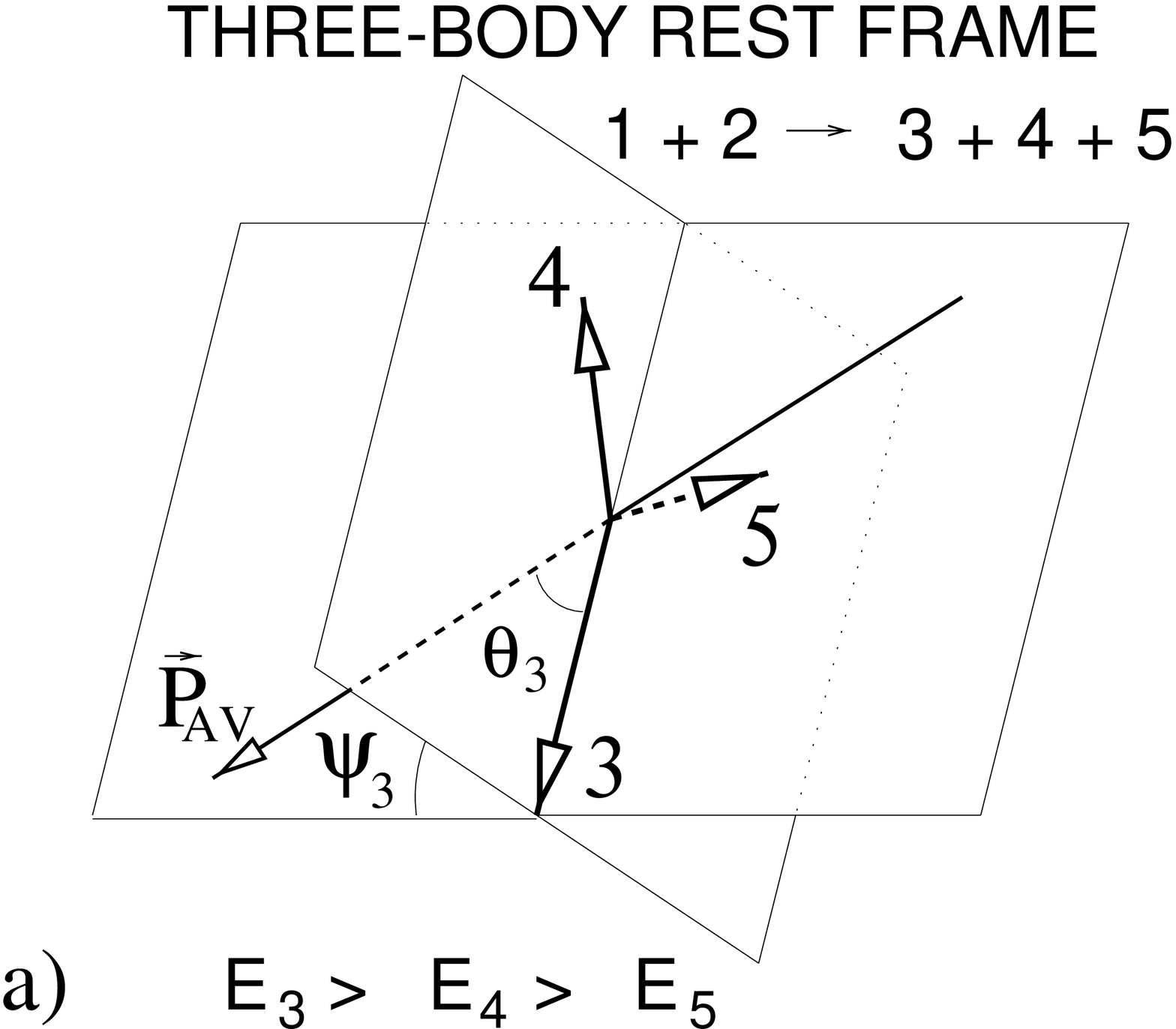}}\makebox[8.cm][l]{\epsfxsize=7.cm\epsfbox{m_sys_z22_py.epsi}}
\caption{a) Illustration of the angles $\theta_3$ and $\psi_3$ for a particular
three jet configuration. b) Three-jet invariant mass distribution.\label{fig:diagram}}
\end{flushright}
\end{figure}

The definition of the angles $\theta_3$ and $\psi_3$ is illustrated in 
figure~\ref{fig:diagram}~a).
$\theta_3$ is the angle between the highest energy jet and the beam direction.
It is similar to the cm scattering angle $\theta^*$ familiar from dijet
studies.  We therefore expect the distribution of $\cos\theta_3$ to have 
approximately a
Rutherford scattering form $\sim (1 - \cos \theta^*)^{-2}$ and to reflect
the spin of the primary exchanged parton.  $\psi_3$ is the angle between the 
plane containing
the three jets and the plane containing the highest energy jet and the beam 
direction.  The coherence property of QCD causes these two planes to tend
to coincide leading to a $\psi_3$ distribution which peaks toward 0 and $\pi$.

\section{Event Selection}

Photoproduction events are selected by restricting the momentum transfer at the
positron vertex to $< 1$GeV and photon-proton cm energies
$W_{\gamma p}$ are selected in the range 134~GeV$ < W_{\gamma p} < $269~GeV.
Jets are found using the {\sc KTCLUS}~\cite{Mike_93} finder in inclusive 
mode~\cite{Ellis_93}.  This is a longitudinally invariant cluster algorithm 
which combines particles with
small relative transverse momenta into jets.  Because it suffers from no 
ambiguities due to overlapping jets it is ideal for the study
of multijet processes.  The jets are required to have transverse energies
$E_T^{\mbox{\scriptsize jet}} > 6$~GeV and pseudorapidities 
$|\eta^{\mbox{\scriptsize jet}}| < 2.5$.  The requirement of high transverse energy
for the jets ensures that the process should be calculable within pQCD.  However
it introduces a bias on angular distributions by throwing away jets
which are produced close to the beam-line.  We make the additional requirements
$M_{3J} > 50$~GeV, $|\cos \theta_3| < 0.8$ and $X_3 < 0.95$ in order relieve the
angular distributions of this effect.

\section{Results}

The three-jet invariant mass distribution is shown in figure~\ref{fig:diagram}~b) and
compared with \oos\ calculations from two groups of
authors, Harris \& Owens~\cite{Harris} and Klasen \& Kramer~\cite{Klasen}.
The pQCD calculations are in relatively good agreement with the data, given
that the calculation is leading order for this process.  The Monte Carlo
models {\sc PYTHIA~5.7}~\cite{PYTHIA} and {\sc HERWIG~5.9}~\cite{HERWIG} contain only 
the two-to-two
matrix elements but are able to produce three jet events through the initial
and final state parton showers.  The $M_{3J}$ distribution predicted by these
models is in agreement in shape with the data although the predicted  
cross section is too low by 30-40\%.

\begin{figure}[htb]
\centering
\epsfxsize=12.cm
\leavevmode
\epsfbox{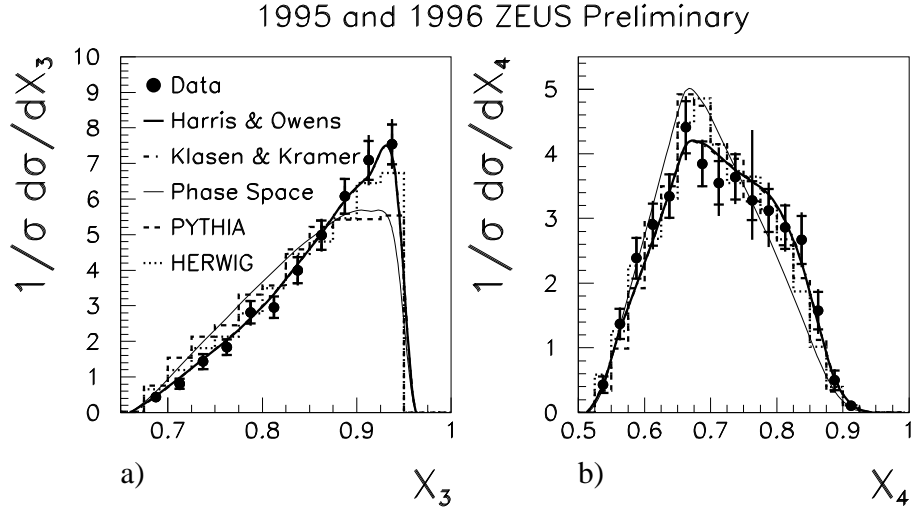}
\caption{a) Highest and b) second-highest energy jet energy fractions.\label{fig:x3x4}}
\end{figure}

The fractions of the available energy taken by the highest and second highest
energy jets are shown in figures~\ref{fig:x3x4} a) and \ref{fig:x3x4} b) 
respectively.
Here the prediction for three jets uniformly distributed in the available
phase space is also shown.  The pQCD calculations (completely overlapping)
and the parton shower models
are all able to describe these energy sharing quantities.  However, notice the
similarity of the measured distributions with the three-body phase space 
distributions.  These distributions have little sensitivity to the pQCD matrix 
element.

\begin{figure}[htb]
\centering
\epsfxsize=12.cm
\leavevmode
\epsfbox{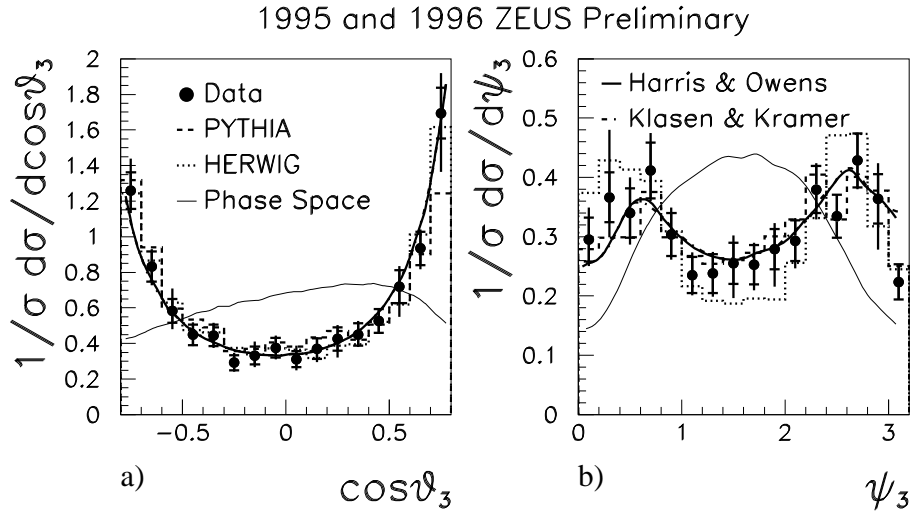}
\caption{a) Scattering angle of highest energy jet and b) angle between three-jet
plane and plane of beam and highest energy jet.\label{fig:cospsi}}
\end{figure}

In figures~\ref{fig:cospsi} a) and \ref{fig:cospsi} b) the 
$\cos \theta_3$ and $\psi_3$ 
distributions are shown.
The measured angular distributions are dramatically different from the 
distributions obtained from phase space.
This shows that these quantities have a high degree of sensitivity to the QCD matrix element.
The $\cos \theta_3$ values are indeed distributed approximately according to 
the Rutherford form.  The \oos\ pQCD and parton shower
calculations, which take into account the dependence of this distribution on
quark and gluon spin, are in excellent agreement with the data.
The jet algorithm and the $E_T^{\mbox{\scriptsize jet}} > 6$~GeV 
requirements have thrown away portions of the phase space near $\psi_3 \sim 0$ 
and $\pi$ as indicated by the shape of the phase space curves.  
Taking this
into account the data indicate a strong tendency for the three-jet plane to
lie near the plane containing the beam and the highest energy jet.
This effect is reproduced in the \oos\ calculations via the QCD
initial state radiation pole.
It is remarkable that the colour coherence phenomenon, implemented as 
angular ordering in parton shower models, allows the leading order Monte 
Carlo programs PYTHIA and HERWIG also to provide a reasonably accurate 
representation of the $\psi_3$ distribution.

\section{Conclusions}

The cross section for photoproduction of three jets with high invariant
mass has been measured.  It is 
described by \oos\ calculations but underestimated 
in leading order parton shower models.  The orientation of the three jets is 
not uniform in the available phase space.  
Both fixed order pQCD calculations which contain the full two-to-three 
matrix elements and Monte Carlo models which produce three jet events
through the parton shower mechanism are able to reproduce the angular
distribution of the jets.  This constitutes a significant advance in
the understanding of multijet photoproduction.

\end{document}